\documentclass[prb,aps,twocolumn,showpacs,superscriptaddress,nofootinbib]{revtex4}
\usepackage{graphicx}  
\usepackage{dcolumn}  
\usepackage{bm}           
\usepackage{amsmath}
\usepackage{epsfig}
\usepackage{indentfirst}
\usepackage{psfrag}
\usepackage{subfigure}
\usepackage{amssymb}
\usepackage{color}
\usepackage[colorlinks,linkcolor=blue,citecolor=blue,urlcolor=blue,hyperindex,driverfallback=dvipdfm]{hyperref}

\def\ii{{\rm i}}  \def\ee{{\rm e}}
\def\rb{{\bf r}}  \def\Rb{{\bf R}}    \def\vb{{\bf v}}
    \def\zz{\hat{\bf z}}  
\def\kb{{\bf k}}    
\def\Qb{{\bf Q}}    
\def\me{m_{\rm e}}  
    
  \def\jb{{\bf j}}
  
\def\lamp{\lambda_{\rm p}}      \def\kp{k_{\rm p}}  \def\wp{\omega_{\rm p}}
\def\pp{{\rm p}}  \def\zb{{\bf z}}  \def\kap{\kappa_{\rm p}}  \def\lame{\lambda_{\rm e}}
  \def\EEb{\vec{\mathcal E}}
\def\AAb{\vec{\mathcal A}}

\begin{document}
\title{Efficient orbital angular momentum transfer between plasmons and free electrons}

\author{Wei Cai}
\affiliation{ICFO-Institut de Ciencies Fotoniques, The Barcelona Institute of Science and Technology, 08860 Castelldefels (Barcelona), Spain}
\affiliation{The Key Laboratory of Weak-Light Nonlinear Photonics, Ministry of Education, School of Physics and TEDA Institute of Applied Physics, Nankai University, Tianjin 300457, China}
\author{Ori Reinhardt}
\affiliation{Faculty of Electrical Engineering and Solid State Institute, Technion, Haifa 32000, Israel}
\author{Ido Kaminer}
\affiliation{Faculty of Electrical Engineering and Solid State Institute, Technion, Haifa 32000, Israel}
\author{F. Javier Garc\'ia de Abajo}
\email[Corresponding author: ]{javier.garciadeabajo@nanophotonics.es}
\affiliation{ICFO-Institut de Ciencies Fotoniques, The Barcelona Institute of Science and Technology, 08860 Castelldefels (Barcelona), Spain}
\affiliation{ICREA - Instituc\'o Catalana de Recerca i Estudis Avan\c cats, Barcelona, Spain}

\date{\today}
\begin{abstract}
Free electrons can efficiently absorb or emit plasmons excited in a thin conductor, giving rise to multiple energy peaks in the transmitted electron spectra separated by multiples of the plasmon energy. When the plasmons are chiral, this can also give rise to transfer of orbital angular momentum (OAM). Here, we show that large amounts of OAM can be efficiently transferred between chiral plasmons supported by a thin film and free electrons traversing it. Under realistic conditions, our predictive simulations reveal efficient absorption of $\ell\gg1$ chiral plasmons of vorticity $m\gg1$, resulting in an OAM transfer $\ell m\hbar\gg\hbar$. Our work supports the use of chiral plasmons sustained by externally illuminated thin films as a way of generating high-vorticity electrons, resulting in a remarkably large fraction of kinetic energy associated with motion along the azimuthal direction, perpendicular to the incident beam.
\end{abstract}
\pacs{42.50.Wk, 45.20.dc, 78.70.g}
\maketitle


\section{\label{sec:level1}Introduction}

Orbital angular momentum (OAM), observed in light\cite{ABS92} and electron\cite{BBS07} waves alike, is associated with staircase wavefronts showing an $\exp(\ii m\varphi)$ dependence on the azimuthal angle $\varphi$ relative to the beam propagation direction, where $m$ is topological charge. Unlike the intrinsic spin angular momentum, which takes finite values ($\pm\hbar/2$ for electrons and $\pm\hbar$ for light), there is no upper limit of OAM. This renders optical OAM particularly attractive for a wide range of applications in optical communications,\cite{MTT07} optical tweezers,\cite{G03} and data storage.\cite{NVG14}

Recently, the production and control of OAM in electron beams has attracted considerable attention,\cite{UT10,VTS10,MAA11,BDN11,BSV12,SSV12,RB13_2,SHH13,paper243,KML16,BIG17,LBT17,SZZ17} particularly with regards to electron-specific phenomena. For example, unlike photons, electron vortices carry charge and magnetic multipoles, which allow probing magnetic transitions.\cite{RB13_2,LBY12} Electron vortices have been also explored as detectors of chirality in crystals\cite{JBA15} and molecules.\cite{paper243} They show even greater potential for optical mode imaging based on the selective excitation of dipolar modes by phase-shaped electron beams.\cite{GBL17}

The interaction between chiral light and electrons opens new directions to explore and exploit vortex phenomena. However, direct interaction between free-space electrons and photons is extremely weak due to energy-momentum mismatch. This is neatly illustrated by the Kapizta-Dirac effect,\cite{KD1933} which consists of elastic electron diffraction by light standing waves, essentially involving virtual processes of photon emission and absorption that modify the phase of the electron wave function in regions of high light intensity. The extremely weak free-space photon-electron interaction and the complexity of the experimental implementation of this effect\cite{FAB01,B07} explain why its demonstration took nearly seven decades since its prediction.\cite{KD1933}

Alternatively, the energy-momentum mismatch can be broken by employing evanescent light\cite{SR1973,H09,H11_2,paper114}
(i.e., light with momentum exceeding the free-space value), as successfully demonstrated to produce multiple photon-electron exchanges in what has been baptized as photon-induced near-field electron microscopy\cite{BFZ09,paper151,PLZ10,FES15,PLQ15,EFS16,KSE16,RB16,VFZ16,MB17,PRY17,paper304} (PINEM). In PINEM, large electron-photon coupling is achieved by temporally synchronizing electron and laser pulses, leading to multiple energy losses and gains. In a related context, surface plasmons
have provided a traditional playground to perform spectroscopy with electron microscopes.\cite{VS1973,B1982,UCT92,KHS00,paper149} As a natural combination of these areas, recent PINEM experiments have revealed insight into the ultrafast dynamics of plasmons evolving in metallic nanowires\cite{PLQ15} and thin films.\cite{paper282} These PINEM studies have focused on energy exchanges, without paying attention to the lateral electron wave function, other than the prediction of efficient diffraction by plasmon gratings.\cite{paper272} However, the exchange of OAM associated with electron-plasmon interactions remains an unexplored area.

In this paper, we show that the interaction between free electrons and chiral plasmons can result in large OAM exchanges. We focus our study on electron Gaussian beams transversing plasmon-supporting thin films, in which plasmons of well-defined chirality are assumed to be optically excited. The OAM acquired by the transmitted electrons is determined by the product of the net number of absorbed or emitted plasmons times $m\hbar$, where $m$ is the topological charge of the plasmons. Coupling to chiral plasmons thus provides a unique tool for producing electron beams with high OAM, which supplements other previously reported approaches\cite{UT10,MAA11,SSV12,BVV13} and presents the advantage of being externally controllable through the applied light intensity and polarization. Our results reveal high transfer efficiencies under realistic experimental conditions. In fact, the electron OAM exchange can be directly observed by energy filtered Fourier-plane electron imaging in existing PINEM setups:\cite{BFZ09,paper151,PLZ10,FES15,PLQ15,EFS16,KSE16,RB16,VFZ16,MB17,PRY17,paper304} as we show below, filtering by different energy windows results in electron {\it donut} beams of different sizes.

\begin{figure}[ht]
\includegraphics[width=80mm,angle=0]{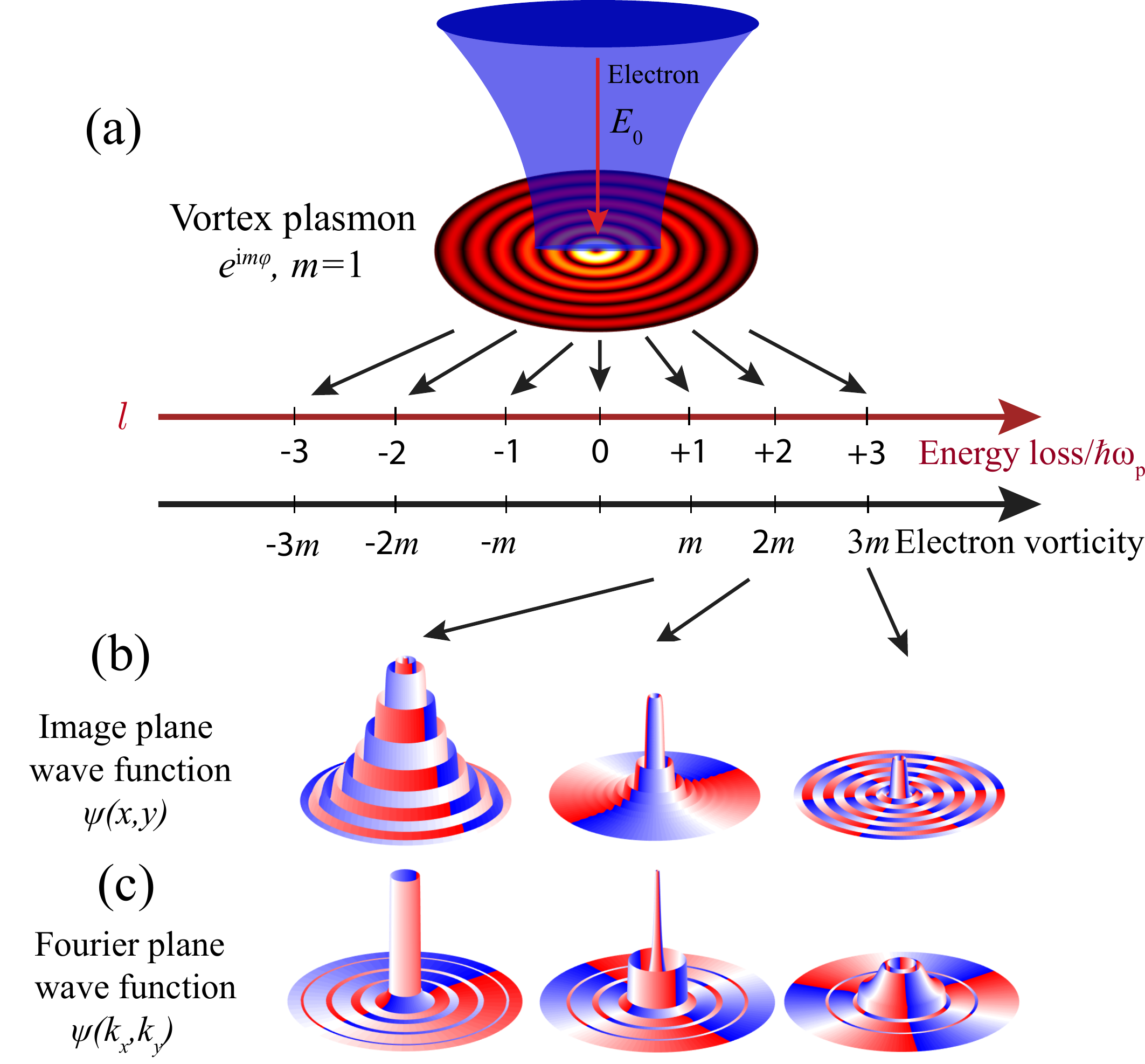}
\caption{{\bf General description of the interaction between free electrons and chiral plasmons.} \textbf{(a)} A Gaussian electron beam interacts with a vortex plasmon confined within an electron-transparent thin metal film. Plasmons of frequency $\wp$ and fixed vorticity (topological charge $m=1$ in this example) are considered. The electron can lose or absorb multiple plasmons, resulting in a transmitted electron spectrum with peaks displaced by integral values of the plasmon energy $\ell\hbar\wp$ relative to the incident electron energy $E_0$. Additionally, the electron changes its OAM by multiples $\ell m\hbar$ of the plasmon OAM $m\hbar$. \textbf{(b,c)} Each transmitted peak $\ell$ at energy $E_0+\ell\hbar\wp$ exhibits a characteristic wave function with $\ell m$ vorticity, evolving from real space in the plane right after interacting with the plasmon (b) to the far-field Fourier plane (c). The color scale indicates the phase of the wave function, whereas ripples in the plots reveal oscillations with periods $\sim\lamp$ and $\sim2\pi/\lamp$ in real and Fourier space, respectively, where $\lamp$ is the plasmon wavelength. In this figure, we take a coupling strength $\eta=2.5$ [see Eq.\ (\ref{etaeta})] and a Gaussian electron beam of $30\,\lamp/2\pi$ full-width-at-half-maximum (FWHM) waist diameter.}\label{fig1} 
\end{figure}

\section{\label{sec:level2.0}Theoretical description}

The system under consideration is schematically depicted in Fig.~\ref{fig1}(a). A coherent electron Gaussian beam of wave function $\psi^{\rm inc}(\textbf{r},t)$ passes through a vortex plasmon field, which can be realized for example by illuminating a thin-film plasmonic vortex lens with circularly polarized light,\cite{KPC10} or by other metasurface designs.\cite{DGB15,SDG15,SKM17} Assuming typical beam divergence angles used in electron microscopes ($\sim$10 mrad), the longitudinal electron momentum is much larger than the transversal component, so we can work in the paraxial approximation. Following a formalism derived elsewhere,\cite{paper151,paper272} we write the incident wave function as
\begin{align}
\psi^{\rm inc}(\rb,t)
\approx\frac{1}{\sqrt{L}}\ee^{\ii(k_0z-E_0t/\hbar)}\phi_0(\rb-\vb t),
\label{psiinc}
\end{align}
where $E_0$, $\hbar k_0$, and $\vb$ are the central energy, momentum, and velocity vector, $L$ is a quantization length along the beam direction, and $\phi_0(\rb)$ describes the Gaussian component of the wave function, which evolves slowly along the beam direction $z$. We further consider the electron-plasmon interaction region to be in the waist of the electron Gaussian beam and to extend over a sufficiently small distance outside the film such that $\phi_0(\rb)$ remains approximately independent of $z$. We thus approximate $\phi_0(\rb)\approx\ee^{-R^2/{\Delta'}^2}/\sqrt{\pi}\Delta'$ in that region, where $\Rb=(x,y)$.
Instead of $\Delta'$, we refer in what follows to the beam intensity full-width-at-half-maximum $\Delta=\sqrt{2\ln2}\Delta'$.

We assume the plasmon to be confined to a thin film (negligible thickness compared with the plasmon wavelength) perpendicular to the beam direction $z$. The electron interacts with the $z$ component of the plasmon electric field,\cite{paper151,paper272} which for fixed topological charge $m$ admits the expression $\mathcal{E}_z(\rb,t)=\mathcal{E}_z(\Rb,z)\ee^{-\ii\wp t}+{\rm c.c.}$, where $\wp$ is the frequency and
\begin{align}
\mathcal{E}_z(\Rb,z)=\mathcal{E}_0\,{\rm{sign}}(z)J_m(\kp R)\ee^{\ii m\varphi}\ee^{-\kap|z|}
\label{Ez}
\end{align}
describes the in-plane variation. This expression describes the plasmon as the combination of two 2D cylindrical evanescent waves of TM polarization (one on either side of the film), which are directly written from an explicit expression given elsewhere [see Eqs.\ (A2) in Ref.\ \onlinecite{paper047}]. Here, $R$ and $\varphi$ are the polar coordinates of $\Rb$, $\mathcal{E}_0$ is an overall constant amplitude, $\kp$ is the plasmon wavenumber (wavelength $\lamp=2\pi/\kp$), $J_m$ is the Bessel function of order $m$, and $\kap=\sqrt{\kp^2-\wp^2/c^2}$. In practice, we consider $\kp\gg\wp/c$ (quasistatic approximation), so we approximate $\kap\approx\kp$. Obviously, the in-plane wave vector $\kp$ and frequency $\wp$ are taken along the plasmon dispersion relation of the film, and in particular, when describing it by means of a local 2D optical conductivity $\sigma(\omega)$, we have\cite{paper235} $\kp\approx\ii\wp/2\pi\sigma(\wp)$, which is the condition derived from the electromagnetic boundary conditions at the film when the electric field is given by Eq.\ (\ref{Ez}). An example of a chiral plasmon is shown in Fig.\ \ref{fig1}(a) (contour plot). Incidentally, this type of chiral plasmon can be excited by shaping the edges of the film with a cicular-saw profile and illuminating with a normal light plane wave. With properly designed film edges, the optical electric field has the form given by Eq.\ (\ref{Ez}) in the central area away from the edges. Additionally, for normally incident light (co-linear with the electron beam), the incident optical field has no component along the $z$ direction.

We now describe the plasmon-electron interaction using a minimal coupling Hamiltonian to obtain the transmitted electron wave function $\psi(\rb,t)$. Right after interacting with the plasmon (i.e., in a plane $z$ close to the sample, but where the plasmonic fields have already decreased to negligible levels), we find that $\psi(\rb,t)$ is given by Eq.\ (\ref{psiinc}) with $\phi_0(\rb-\vb t)$ replaced by\cite{paper151,paper272} (see details in the Appendix)
\begin{align}
\phi(\rb,t)=\phi_0(\rb-\vb t)\sum_{\ell=-\infty}^\infty\ee^{\ii\ell\wp(z/v-t)}f_\ell(\Rb)
\label{eqn4}
\end{align}
where
\begin{subequations}
\begin{align}
&f_\ell(\Rb)=\ee^{\ii\ell\arg\{-\beta\}}\,J_\ell(2|\beta|), \label{fbetaa}\\
&\beta=\frac{e\gamma}{\hbar\wp}\int_{-\infty}^\infty dz\, \mathcal{E}_z(\Rb,z)\,\ee^{-\ii\wp z/v}, \label{fbetab}
\end{align}
\label{fbeta}
\end{subequations} \noindent
$v$ is the electron velocity, $\gamma=1/\sqrt{1-v^2/c^2}$, and the $\Rb$ dependence of $f_\ell(\Rb)$ is inherited from $\mathcal{E}_z(\Rb,z)$. The transmitted electron wave function consists of wave packets of energy $E_0+\ell\hbar\wp$ and momentum $\hbar(k_0+\ell\wp/c)$, labeled by the net number of exchanged plasmons $\ell$ [see Eq.\ (\ref{eqn4})]. Inserting Eq.\ (\ref{Ez}) into Eqs.\ (\ref{fbeta}), we readily obtain
\begin{align}
f_\ell(\Rb)=\ii^\ell\,\ee^{\ii\ell m\varphi}\;J_\ell\left[2\eta\,J_m(\kp R)\right],
\label{near}
\end{align}
where
\begin{align}
\eta=2g\Omega/(1+\Omega^2) \label{etaeta}
\end{align}
is a dimensionless electron-plasmon coupling-strength parameter, defined in terms of the normalized plasmon amplitude $g=e\mathcal{E}_0\lamp\gamma/2\pi\hbar\wp$ and $\Omega=\wp\lambda_\pp/2\pi v$ (the number of optical cycles required by the electron to move along a distance $\lamp$). We note that the electron velocity enters Eq.\ (\ref{etaeta}) (and therefore the interaction strength) only through $\Omega$, and that a maximum interaction strength $\eta$ is achieved when $\Omega=1$, corresponding to the condition that the electron takes an optical cycle ($2\pi/\wp$) to move along the distance of a plasmon wavelength $\lamp$. This seems to be the optimum condition for compensating the sign cancellation in the interaction with the plasmon field on either side of the film produced by the factor ${\rm sign}(z)$ in Eq.\ (\ref{Ez}), keeping also in mind that the plasmon field flips sign twice during one optical cycle.

\begin{figure}[ht]
\includegraphics[width=70mm,angle=0]{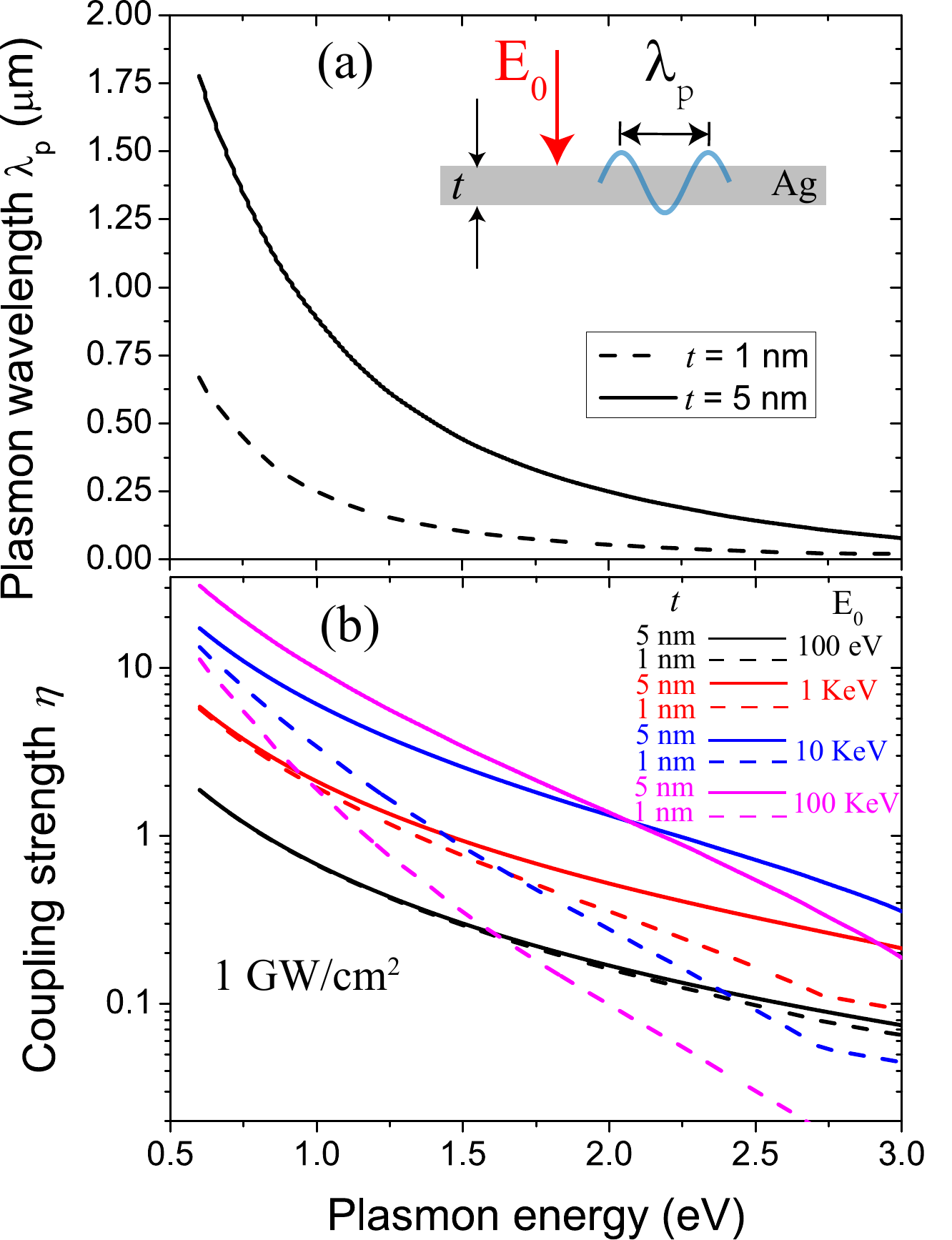}
\caption{{\bf Electron-plasmon coupling strength in thin metal films.} {\bf (a)} Plasmon wavelength as a function of energy for silver films of 1\,nm and 5\,nm thickness. {\bf (b)} Coupling strength $\eta$ [see Eq.\ (\ref{etaeta})] for 1\,nm (broken curves) and 5\,nm (solid curves) silver films and various electron energies as indicated in the legend.} \label{fig2}
\end{figure}

The coupling strength $\eta$ plays a central role in this study. We plot it for thin silver films in Fig.\ \ref{fig2}, where we show that the plasmon wavelength is small compared with the light wavelength above 0.5\,eV energy, thus validating the use of the quasistatic approximation. We find that $\eta$ shows a strong dependence on electron energy and
is generally larger for longer plasmon wavelengths. In what follows, we focus on the attainable range $\eta=0-20$.

In experiment, one can prepare the electron optics of a microscope to directly record the electron intensity $|\psi(\Rb,z)|^2$ at an image plane $z$ right after interaction with the plasmon. An example of the expected result is illustrated in Fig.\ \ref{fig1}(b), showing characteristic ripples with $\sim\lamp$ period corresponding to the spatially modulated amplitude of the plasmon vortex field. Additionally, one can record the transmitted electrons in the far field (i.e., the Fourier plane), also illustrated in Fig.\ \ref{fig1}(c). We apply scalar diffraction theory\cite{J99} to calculate the Fourier-plane amplitude $f_{\ell,\Qb}$ from the image-plane amplitude as (see Appendix)
\begin{align}
f_{\ell,\Qb}=C\,\ee^{\ii\ell m\varphi_\Qb}\!\!
\int_0^\infty\!\!\!\!\! RdR\,J_\ell\left[2\eta\,J_m(\kp R)\right]
J_{\ell m}(QR),
\label{far}
\end{align}
where $\Qb$ is the transversal electron wave vector (in the $x$-$y$ plane) and $C=2\pi\ii^{\ell(1-m)}$.

We are interested in the image- and Fourier-plane intensities associated with different numbers of exchanges $\ell$ between the electron and a plasmon of vorticity $m$, as illustrated in Fig.\ \ref{fig1}. These two planes directly correspond to two avenues for observing and utilizing the resulting electron vortex beams  using the transition electron microscope (TEM) in imaging and diffraction modes. We also consider the transmitted electron currents along the initial beam direction $z$ and along the transversal azimuthal direction. Inserting the wave function associated with each channel $\ell$ into the expression $\jb={\rm Re}\{-\ii\hbar\psi^*\nabla\psi/m_e\gamma\}$ for the current density, making the substitution $\nabla\rightarrow\ii k_0\hat{\zb}+\ell m\hat{\bf{\varphi}}/R$, and integrating over $\Rb$, the current components transmitted from the incident Gaussian beam reduce to
\begin{align}
\begin{bmatrix} I_z \\ I_\varphi \end{bmatrix}
=\frac{2I_{\rm{inc}}}{\pi{\Delta'}^2}\int_0^{\infty}\!\!\!dR\,|f_{\ell}(R)|^2\ee^{-2R^2/{\Delta'}^2}
\begin{bmatrix} 2\pi R \\ \ell m\lame \end{bmatrix},
\label{current}
\end{align}
where $I_{\rm{inc}}$ is the incident electron intensity and $\lame$ is the electron wavelength.

\section{\label{sec:level2}Results and discussion}

The process of plasmon absorption or creation by the electron must conserve OAM. Each plasmon exchange therefore involves an OAM $m\hbar$, so that a net number of plasmon exchanges $\ell$ results in a transfer of OAM given by $\ell m\hbar$. Figure~\ref{fig1} illustrates this for $m=1$: besides the noted ripples in the intensity at the image and Fourier planes, we superimpose a color map showing the phase of the wave function, as calculated from Eqs.~(\ref{near}) and (\ref{far}), revealing a modulation along the azimuthal angle $\varphi$ with $\ell m$ phase jumps from $-\pi$ to $\pi$, arising from the $\ee^{i\ell m\varphi}$ factors in those equations.

\begin{figure}[ht]
\includegraphics[width=80mm,angle=0]{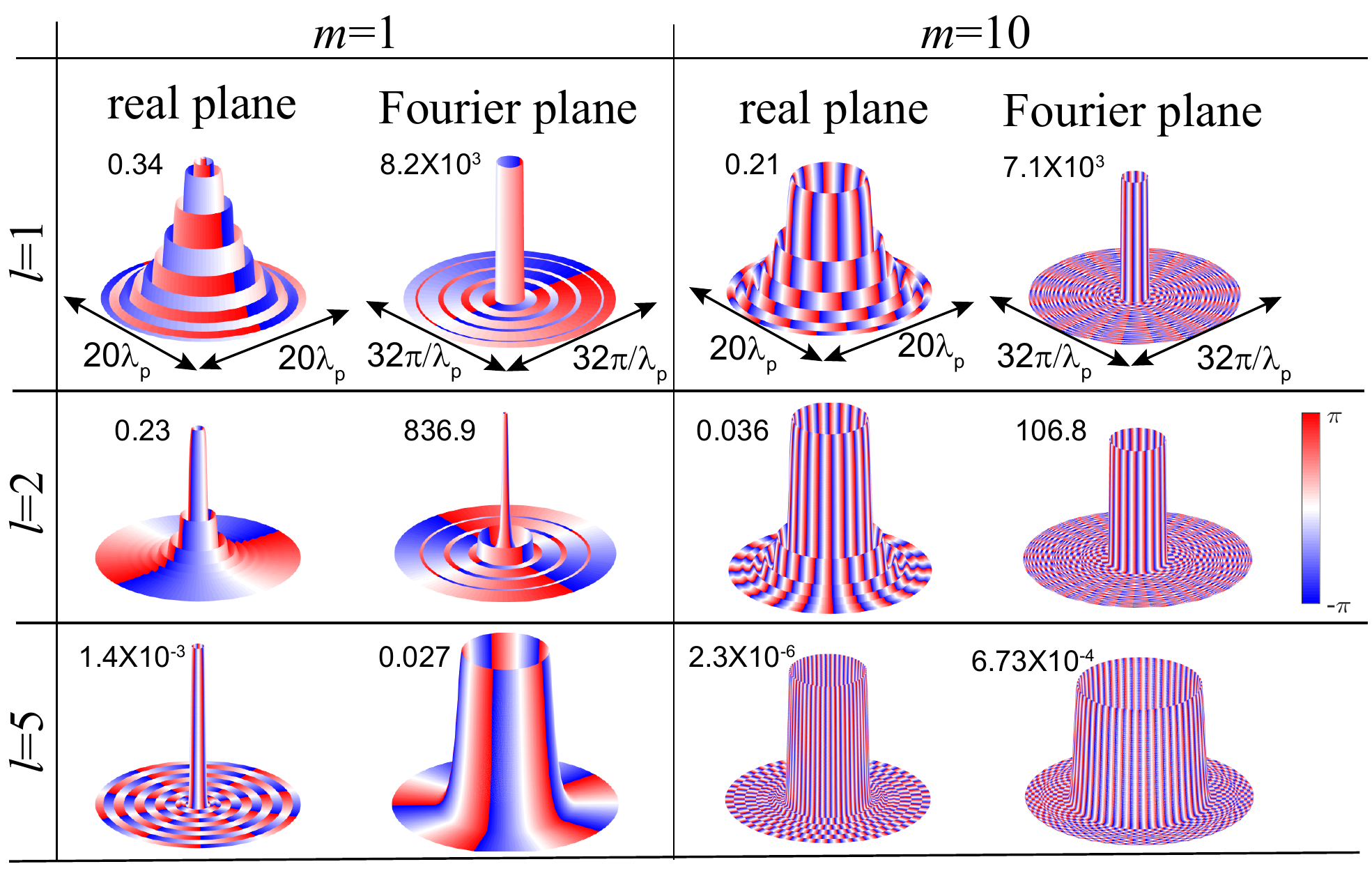}
\caption{{\bf Toward extreme angular momentum transfers during plasmon-electron interaction.}  We show the transmitted electron wave function in real and Fourier space following a net exchange of $\ell=1$, 2, and 5 plasmons of vorticity $m=$1 and 10. The incident Gaussian electron beam has a FWHM $\Delta=30\,\lamp/2\pi$. The coupling strength is set to $\eta=2.5$ [see Eq.\ (\ref{etaeta}) and Fig.\ \ref{fig2}]. Numerical labels in each plot indicate the relative intensities.} \label{fig3}
\end{figure}

A more detailed analysis is presented in Fig.~\ref{fig3} for plasmons of vorticity $m=1$ and 10. We consider a net number of plasmon exchanges $\ell=1$, 2, and 5. Remarkably large values of the OAM can be reached, observed through the fast phase oscillations along the azimuthal direction for $m=10$ and $\ell=5$, leading to an OAM exchange as large as 50$\hbar$, with probability around $4\%$ and reaching unity-order values for stronger field or lower values of OAM. We note that for fixed $m$ the wave function intensity decreases with increasing transfer order $\ell$ both in real and Fourier spaces. This suggests that increasing $m$ is a more efficient way of reaching high OAM transfers than increasing $\ell$, as neatly illustrated by comparing the results obtained for $(\ell,m)=(1,10)$ and $(\ell,m)=(5,1)$, although there is a trade-off between $m$ and $\Delta$: larger $m$ inevitably involves a more widespread plasmon intensity pattern (main lobe radius $\sim m\lamp/2\pi$), which requires a wider $\Delta$ and transverse coherence of the electron beam. With practical implementations in mind, we note that controlling $m$ can be done through the external laser illumination or the sample geometry; however, controlling $\ell$ is achieved by energy filtering of the electron beam, either before the interaction with a target or as post-selection after the desired interaction.

\begin{figure}[ht]
\includegraphics[width=80mm,angle=0]{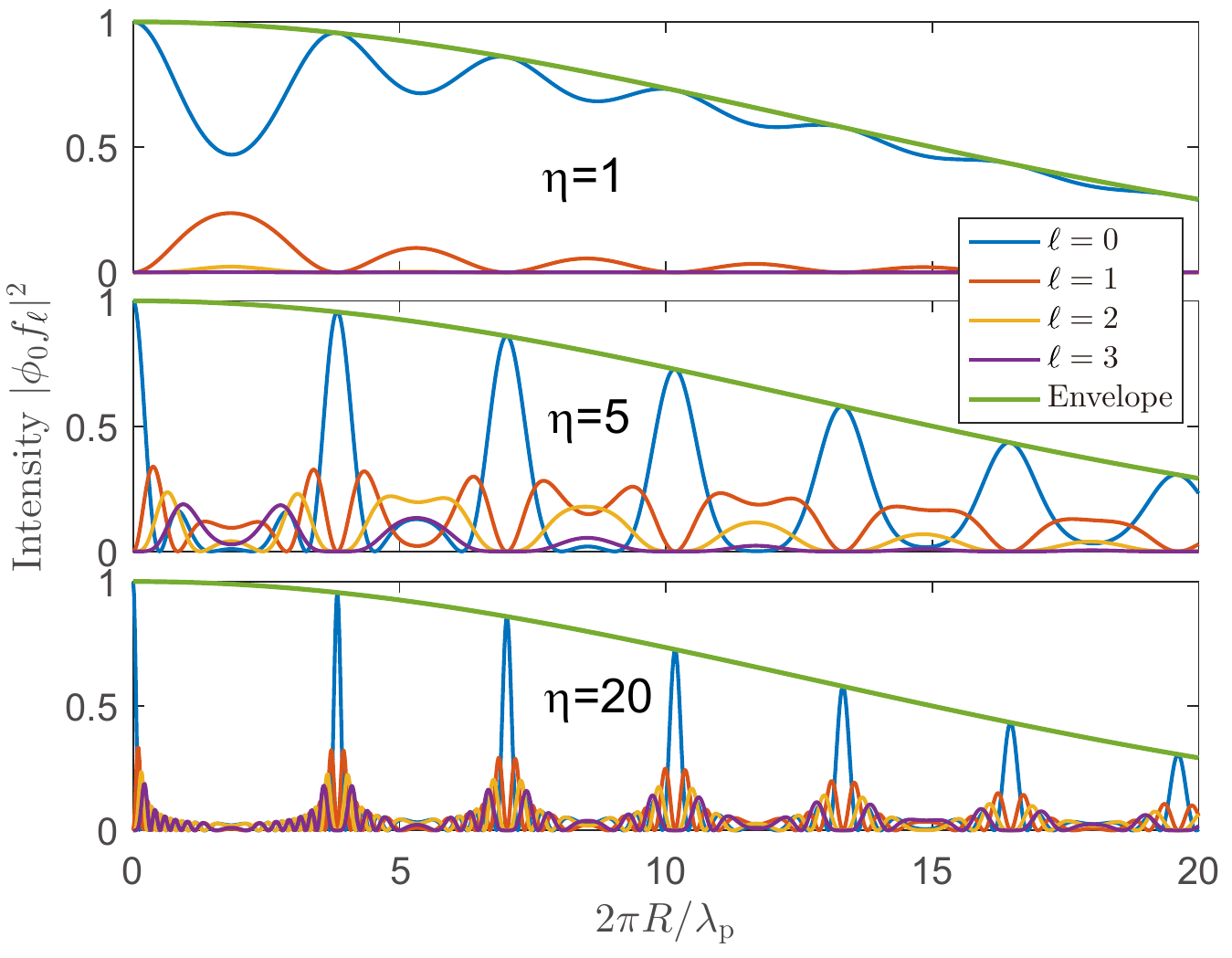}
\caption{{\bf Geometrical effects in the plasmon-transfer efficiency.} We plot the contribution to the electron probability for absorbing $\ell=0-3$ plasmons of vorticity $m=1$ as a function of the lateral position $R$ relative to the plasmon vortex center for different coupling strengths $\eta$ and fixed Gaussian electron beam FWHM $\Delta=30\,\lamp/2\pi$. The probability is normalized to the value at the electron beam center.}
\label{fig4}
\end{figure}

Like any vortex wave, chiral plasmons have vanishing intensity at the origin [see Fig.~\ref{fig1}(a)], which is precisely where the electron Gaussian profile finds its maximum. It is therefore useful to study the dependence of the OAM transfer efficiency on the relative values of the electron Gaussian width and the plasmon wavelength. In Fig.\ \ref{fig4}, we show the probability that an electron absorbs or emits different numbers $\ell$ of plasmons of vorticity $m=1$ as a function of lateral position relative to the center of the plasmon vortex. The coupling efficiency decreases with the displacement, although it exhibits periodic modulations that reflect constructive/destructive superpositions of different spatial regions contributing to the overall transition amplitudes. These modulations become sharper and deeper when we increase the coupling strength $\eta$, as we illustrate for $\eta=$1, 5, and 20, with the maxima signaled by the zeros of $J_m(2\pi R/\lamp)$ for all $\eta$'s.

\begin{figure}[ht]
\includegraphics[width=80mm,angle=0]{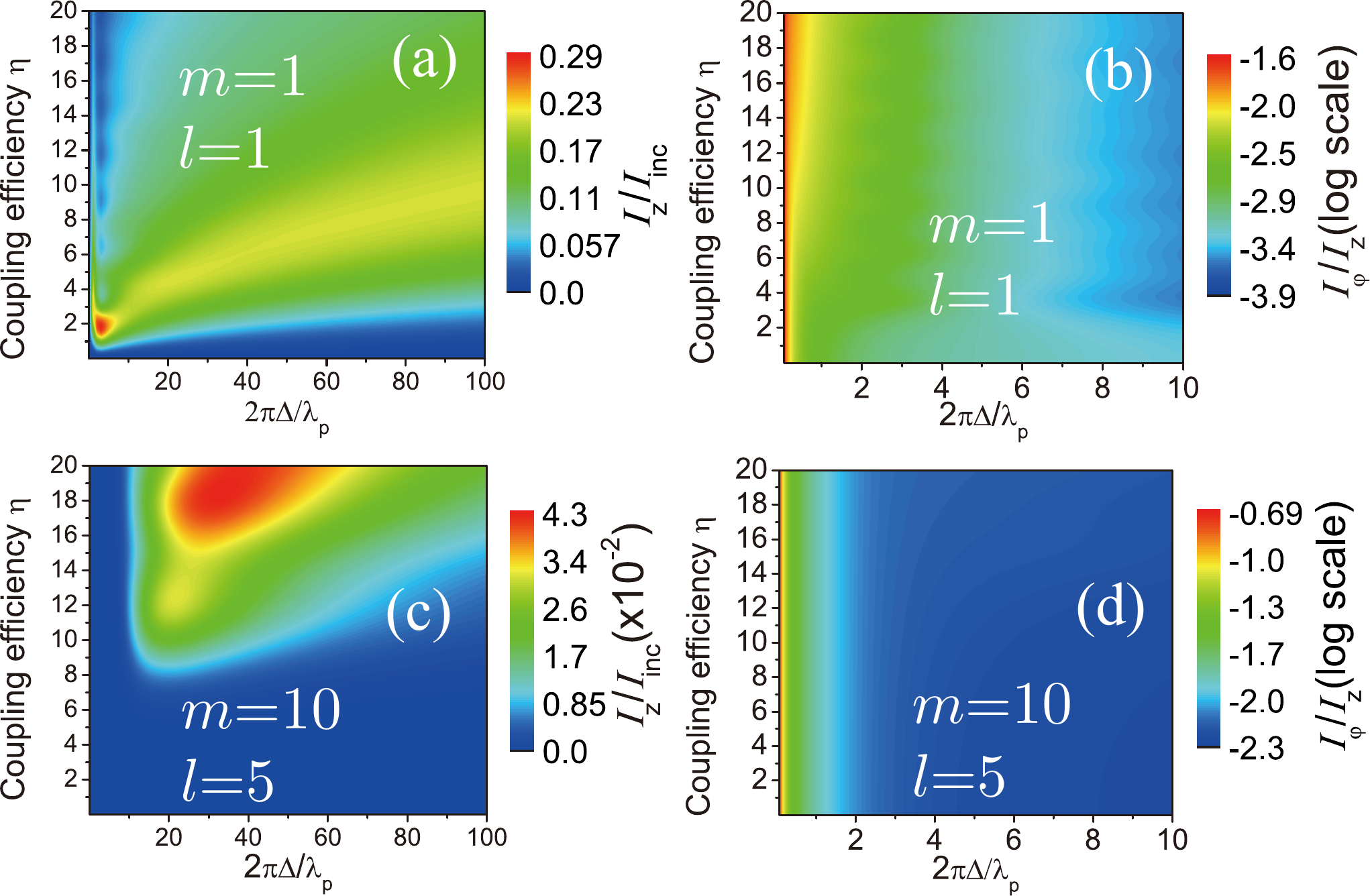}
\caption{{\bf Azimuthal electron current generated by interaction with chiral plasmons.} {\bf (a,c)} Fraction of electron current along the beam direction $I_z/I_{\rm inc}$ after interaction with plasmons of vorticities $m=1$ (a) and $m=10$ (c) with $\ell=1$ and $\ell=5$ net exchanges, respectively. {\bf (b,d)} Fraction of azimuthal current $I_\varphi/I_z$ relative to the scattered current under the conditions of (a,c). Results are plotted as a function of Gaussian electron beam FWHM $\Delta$ and coupling efficiency $\eta$. A ratio of electron-to-plasmon wavelengths $\lambda_{\rm e}/\lamp=1.2\times10^{-3}$ is assumed.}
\label{fig5}
\end{figure}

In order to enable realistic applications of electron-plasmon OAM transfer, it is important to estimate the resulting azimuthal electron current that can be generated by the processes under consideration. In Fig.~\ref{fig5}, we plot the azimuthal current as a function of coupling strength $\eta$ and Gaussian electron beam FWHM $\Delta$ for interaction with $m=1$ (top) and $m=10$ (bottom) plasmons, as obtained from Eq.~(\ref{current}). Obviously, the azimuthal current has a strong dependence on the exchange order $\ell$. 
As we increase the coupling strength $\eta$, the $z$ and $\varphi$ components of the current gradually grow and eventually saturate. Remarkably, the azimuthal current reaches $\sim20\%$ of the total scattered current for the assumed electron-to-plasmon wavelength ratio $\lambda_{\rm e}/\lamp=1.2\times10^{-3}$, which could be obtained for instance with low energy electrons ($E_0=100\,$eV) and well-confined plasmons ($\lamp=100\,$nm). Additionally, the azimuthal component gains weight with increasing $\ell m$.

\section{\label{sec:level6}Concluding remarks}

Looking toward an experimental observation of our work, one can follow recently demonstrated approaches for generating plasmonic vortices.\cite{KPC10, GDB14,SKM17,OCT18} In these and other studies, plasmon fields are probed point-by-point using various techniques (SNOM, PEEM, STEM-EELS). Our prediction is unique in using the coherent electron-plasmon interaction, which depends on the quantum wave function of the electron, which has become possible thanks to recent advances in ultrafast electron microscopy.\cite{VFZ16,FES15,paper304,VMB18} Note that our approach of using the coherent electron-photon interaction for imbuing the electron with OAM is unique in comparison to current methods for shaping electrons with OAM.\cite{UT10,VTS10,MAA11} In most of the current literature the electron is treated as an analog of a photon and is shaped by phase masks similar to the ones used to shape light. In contrast, our approach exploits the charged nature of the electron and its ability to interact with light in a quantum mechanical manner,\cite{paper151,PLZ10,FES15,paper272} which brings new ways to control OAM exchange.

In conclusion, we have shown that OAM can be efficiently transferred from plasmons to electrons when an electron Gaussian beam transverses a vortex plasmon field supported by a thin film. For electrons that have absorbed $\ell$ plasmons of vorticity $m$, the transfered OAM is given by $\ell m\hbar$, which can reach high values with attainable coupling strengths. The latter depend strongly on the lateral size of the electron beam relative to the plasmon wavelength, as well as on the electron velocity, as we illustrate for thin silver films. We predict that in practice a fraction as high as 4\,\% of the incident electrons can be transmitted with high vorticity ($\ell m\sim$50) under attainable conditions. Our findings support the use of electron-plasmon coupling in PINEM as a way to prepare highly chiral electron beams, which adds up to the existing suite of techniques relying on direct phase imprinting,\cite{UT10} diffraction by holographic masks,\cite{MAA11} mode conversion,\cite{SSV12} and interaction with pseudo magnetic monopoles.\cite{BVV13} We remark that our proposed approach presents the advantage of being dynamically tunable through the degree of polarization and intensity of the incident light. Despite the technical difficulties, a first demonstration of OAM transfer between chiral plasmons and electron beams has been recently demonstrated.\cite{Vanacore} Ultimately, plasmon-electron OAM transfer is a tool to shape the phase of the electron wave function, which we have explored for Bessel waves. This scheme can be directly extrapolated to other types of electron wave functions, such as Airy beams\cite{VLL13} using Airy plasmons.\cite{SC10,MKJ11} Other kinds of evanescent waves could equally be exploited instead of plasmons, such as phonon polaritons in 2D materials,\cite{DFM14,paper283} which could provide a flexible range of mode wavelengths and coupling strengths.

\section*{Acknowledgments}

We thank F. Carbone, G. M. Vanacore, I. Madan, and G. Berruto for helpful and enjoyable discussions. This work has been supported in part by the Program for Changjiang Scholars and Innovative Research Team in University (IRT13\_R29), the National key R\&D Program of China (2017YFA0305100, 2017YFA0303800), the National Natural Science Foundation of China (61775106, 11374006, 11774185, 11504184), the 111 Project (B07013), the European Research Council (Marie Curie 328853-MC-BSiCS),  the Spanish MINECO (MAT2017-88492-R and SEV2015-0522), CERCA, Fundaci\'o Privada Cellex, and AGAUR (2014 SGR 1400) programs. I.K. acknowledges a Fellowship from the Azrieli Foundation.

\appendix

\section*{Appendix}

\section*{Derivation of Eqs.\ (\ref{eqn4}) and (\ref{fbeta})}

For the sake of completeness, we provide a derivation of Eq.\ (\ref{eqn4}) of the main paper following previous formulations \cite{paper151,PLZ10,paper272}. In particular, we sketch a recent derivation presented in a recent publication \cite{VMB18}. The interaction of the electron beam with an optical field can be described through the Schr\"odinger equation $(H_0+H_1)\psi=\ii\hbar\partial\psi/\partial t$, where $\psi(\rb,t)$ is the electron wave function, $H_0$ is the noninteracting Hamiltonian, and $H_1=(-\ii e\hbar/\me c)\AAb(\rb,t)\cdot\nabla$ accounts for electron-light interaction through the vector potential $\AAb(\rb,t)$. The electron wave function is made of components $\ee^{\ii(\kb\cdot\rb-E_\kb t/\hbar)}$ of well-defined momentum $\hbar\kb$, centered around a central value $\kb_0=\hbar^{-1}\sqrt{(2\me E_0)(1+E_0/2\me c^2)}\zz$, where $E_0$ is the nominal electron kinetic energy. The energy of each component $\kb$ differs slightly from $E_0$, so it can be expanded as $E_\kb\approx E_0+\hbar\vb\cdot(\kb-\kb_0)$, where $\vb=(\hbar\kb_0/\me)/(1+E_0/\me c^2)$ is the nominal electron velocity. This approximation, which is valid for small momentum spread (i.e., $|\kb-\kb_0|\ll k_0$), allows us to approximate $H_0\approx E_0-\hbar\vb\cdot(\ii\nabla+\kb_0)$. We also approximate $\nabla\approx\ii\kb_0$ in $H_1$ (i.e., $\nabla$ is dominated by the fast variation of the electron wave function along the beam direction). At this point, it is convenient to write $\psi(\rb,t)=\ee^{\ii(\kb_0\cdot\rb-E_0t/\hbar)}\phi(\rb,t)$, which upon insertion into Schr\"odinger equation leads to
\begin{align}
(\vb\cdot\nabla+\partial/\partial t)\,\phi=\frac{-\ii e\gamma\vb}{\hbar c}\cdot\AAb\,\phi,
\label{sch1}
\end{align}
where $\gamma=1/\sqrt{1-v^2/c^2}$. Equation\ (\ref{sch1}) has the solution
\begin{align}
&\phi(\rb,t)= \label{solution0}\\
&\phi_0(\rb-\vb t)\,\exp\left[\frac{-\ii e\gamma\vb}{\hbar c}\cdot\int_{-\infty}^t dt'\,\AAb(\rb+\vb t'-\vb t,t')\right], \nonumber
\end{align}
where $\phi_0(\rb-\vb t)$ is the incident electron wave function. For monochromatic light of frequency $\wp$ and electric field amplitude $\EEb(\rb)$, we can write $\AAb(\rb,t)=(-\ii c/\wp)\EEb(\rb)\ee^{-\ii\wp t}+{\rm c.c.}$, which upon insertion into Eq.\ (\ref{solution0}), for a time and position well beyond the region of interaction with the plasmons, yields
\begin{align}
\phi(\rb,t)=\phi_0(\rb-\vb t)\,\exp\left[-\ee^{\ii\wp(z/v-t)}\beta+{\rm c.c.}\right]
\nonumber
\end{align}
with $\beta(\rb)$ given by Eq.\ (4b) of the main paper. We now use the Jacobi-Anger expansion $\ee^{\ii u\sin\varphi}=\sum_{\ell=-\infty}^\infty J_\ell(u)\ee^{\ii\ell\varphi}$ (see Eq.\ (9.1.41) of Ref.\ \onlinecite{AS1972}), taking $|u|=2|\beta|$ and $\varphi=\arg\{-\beta\}$, to write Eq.\ (\ref{eqn4}) of the main paper with $f_\ell(\Rb)$ defined as in Eq.\ (fbetaa).

\section*{Derivation of Eq.\ (\ref{far})}

We start from Eqs.\ (\ref{eqn4}) and (\ref{fbeta}) of the main paper, which describe the transmitted electron wave function right after interaction with the plasmon, and use scalar diffraction theory to construct the far-field wave function (i.e., in the Fourier plane). The transmitted electron wave-function component corresponding to a net number of plasmon exchanges $\ell$ can be Fourier-transformed and expressed as a combination of components $\psi_{\ell,\Qb}(\rb)$ with well defined transversal wave vector $\Qb$, satisfying the Helmhotz equation $(\nabla^2+k_\ell^2)\psi_{\ell,\Qb}=0$, where $k_\ell=k_0+\ell\wp/v$ is the corresponding electron wave number. These components then have a spatial dependence $\psi_{\ell,\Qb}(\rb)\propto\ee^{\ii\Qb\cdot\Rb+\ii k_{z,\ell,\Qb} z}$, where $k_{z,\ell,\Qb}=\sqrt{k_\ell^2-Q^2}$. From Eq.\ (\ref{eqn4}) of the main paper, we can now write the transmitted wave function as
\begin{align}
&\psi^{\rm trans}(\rb,t)\approx \nonumber\\
&\frac{1}{\sqrt{AL}}\sum_\ell\ee^{-\ii E_0t/\hbar-\ii\ell\wp t}\int \frac{d^2\Qb}{(2\pi)^2}\;\ee^{\ii\Qb\cdot\Rb+\ii k_{z,\ell,\Qb} z}\;
f_{\ell,\Qb}, \nonumber
\end{align}
where
\begin{align}
f_{\ell,\Qb}=\sqrt{A}\int d^2\Rb\; \ee^{-\ii\Qb\cdot\Rb} \phi_0(\Rb,0)f_\ell(\Rb),
\label{fff}
\end{align}
and we have introduced a normalization area $A$ in the plane normal to the beam direction. We note that the amplitude $f_{\ell,\Qb}$ depends in general on the transversal profile of the incident electron wave function at the position and time of interaction with the plasmons [i.e., $\phi_0(\rb-\vb t)$ for $z=0$ and $t=0$]. Assuming an incident electron plane wave [$\phi_0(\Rb,0)=1/\sqrt{A}$], substituting $f_\ell(\Rb)$ in Eq.\ (\ref{fff}) by Eq.\ (5) of the main paper, and using the identity
\begin{align}
\int_0^{2\pi}\!\!\! d\varphi\,\ee^{\pm\ii\Qb\cdot\Rb}\;\ee^{\ii n\varphi}=2\pi\ii^{\pm n}\;\ee^{\ii n\varphi_\Qb}\;J_n(QR),
\nonumber
\end{align}
where the integral is over the azimuthal angle of $\Rb$ and $\varphi_\Qb$ is the azimuthal angle of $\Qb$, we readily find Eq.\ (\ref{far}) of the main paper for the far-field Fourier-space amplitude $f_{\ell,\Qb}$.

\section*{Derivation of Eq.\ (\ref{current})}

We start from Eqs.\ (\ref{psiinc}) and (\ref{eqn4}) of the main paper, which give the incident and transmitted electron wave functions $\psi$, and use them to calculate their associated electron current densities as $\jb={\rm Im}\{\hbar\psi^*\nabla\psi/\me\gamma\}$. The current transmitted along each channel $\ell$ admits the substitution $\nabla\rightarrow\ii k_0\hat{\zb}+\ell m\hat{\bf{\varphi}}/R$, leading to the components
\begin{align}
&j_z=\frac{\hbar k_0}{\me L}|\phi_0 f_\ell(\Rb)|^2, \nonumber \\
&j_\varphi=\frac{\ell m\hbar}{\me LR}|\phi_0 f_\ell(\Rb)|^2
\nonumber
\end{align}
along the beam and azimuthal directions, respectively. Integrating these expressions over the transversal coordinate $\Rb$, we find the corresponding currents given by Eq.\ (\ref{current}) of the main paper, in which $I_{\rm{inc}}=\hbar k_0/2\me L$ is the incident electron current, and we approximate the incident beam profile by a Gaussian $\phi_0(\rb)\approx\ee^{-R^2/{\Delta'}^2}/\sqrt{\pi}\Delta'$ at the position of the plasmon-supporting film. It is worth noting that the normalized current along $\varphi$ directions is proportional to the electron wavelength $\lame$.


\end{document}